\documentstyle[12pt,oldlfont,titlepage,a4]{article}
\newcommand{\be}{\begin{equation}}
\newcommand{\ee}{\end{equation}}
\newcommand{\beqs}{\begin{eqnarray*}}
\newcommand{\lb}{\label}
\newcommand{\eeqs}{\end{eqnarray*}}
\newcommand{\bi}{\begin{itemize}}
\newcommand{\ei}{\end{itemize}}

\newcommand{\bt}{\beta}

\newcommand{\lan}{\langle}
\newcommand{\ran}{\rangle}
\newcommand{\hs}{{\cal H}(\{S_i\})}

\newcommand{\spv}{\sum_{<i,j>}}
\newcommand{\si}{\{S_i\}}

\newcommand{\gammaijpar}{\gamma_{ij}^{\parallel}}
\newcommand{\gammaijnotpar}{\gamma_{ij}^{\not\parallel}}

\begin{document}
\begin{center}
{\LARGE \bf  Percolation and Cluster Formalism in \\

\vspace{0.5cm}

Continuous Spin Systems}

\vspace{2cm}

{\Large \sl Mario Nicodemi}

\vspace{2cm}

{\em
Dipartimento di Scienze Fisiche, Universit\`a di Napoli, \\
Mostra d'Oltremare Pad.19, I-80125 Napoli, Italy \\
Sezione INFM and INFN di Napoli

\vspace{0.5cm}

e.mail: nicodemim@axpna1.na.infn.it}


\vspace{2cm}

{\bf Abstract}

\end{center}

The generalization of Kasteleyn and Fortuin clusters formalism is introduced 
in $XY$ (or more generally $O(n)$) models. Clusters geometrical structure 
may be linked to spin physical properties as correlation functions. 
To investigate percolative characteristics,
the new cluster definition is analytically explored in one dimension and with 
Monte Carlo simulations in 2D and 3D frustrated and unfrustrated $n$-clock 
models. 

\pagebreak

\baselineskip=0.5cm

\section{Introduction}

The idea to describe long range correlations and coherency in spin 
systems from a geometrical point of view dates thirty years back. 
In the late sixties this project was accomplished for Ising models by 
Kasteleyn and Fortuin (KF), who developed a method to give intrinsic 
definitions of clusters of spins which might describe with their percolation 
characteristics the structure of correlations patterns.
Cluster formalism and percolation tools \cite{KF,CK}, 
have proven extremely useful in the understanding of critical phenomena of 
Ising models \cite{droplets,Wu}. Among the many results, very interesting is 
the discovery of the possibility to describe spin correlations trough 
percolative connectivity functions and the consequent link between 
thermodynamic critical behaviors and cluster fractal structures 
(see \cite{89Co}). 
The individuation of ``physical" clusters of spin introduced within 
this approach, has been also successfully exploited by Swendsen and Wang 
(SW) \cite{SW,Wolff} to develop fast Monte Carlo (MC) dynamics, for 
unfrustrated Ising models, based on 
cluster update and later on to drastically improve simulations 
in frustrated systems too \cite{BADK,noistessi}. 

Nevertheless, the discussions about the extension of cluster formalism 
and percolation concepts to continuous spin systems as $XY$ or $O(n)$ 
models is still open, and the equivalent of KF cluster in such systems 
is not known. Wolff \cite{Wolff} has proposed a cluster definition in 
$XY$ models based on a smart application of KF rule 
to spin projections along random directions. The clusters so introduced 
have proven to have a percolative critical temperature exactly equal to 
the thermodynamic one in unfrustrated $XY$ models, also if their 
ultimately connections with spin properties is not understood \cite{Henley}.
One of the successes of KF approach consists in the clarification of 
the links between the clusters and the physics of the spins.

In this paper we try to address a possible generalization to $XY$ and $O(n)$ 
models of KF approach to Ising like systems.
This extension leads to new cluster definitions. 
In the spirit of Kasteleyn-Fortuin and Coniglio-Klein (CK) works, we try to 
focus the relations between clusters and spins, disregarding, at 
the moment, applications to efficient MC algorithms. 
Specifically we try to introduce concepts and tools to manipulate the 
the structure of such clusters
in this larger context, as done by KF and CK in Ising like systems.
In ferromagnetic models, the new clusters 
generally individuate regions of statistically coherent spins, i.e. 
almost parallel spins, and describe the physics of such aggregates. 
For sake of clarity, before passing to such a generalization (presented 
in section 3 and 4), 
in section 2 a fast outlook to KF original approach to Ising systems 
is given, to define notations and 
concepts used in what follows. Later, the properties of these new clusters 
are analytically studied in one dimension and via MC simulations in
$n$-clock models in 2D and 3D, where the thermodynamic transitions have 
different properties. 

\section{Cluster formalism for Ising spin Hamiltonians}

Let us consider an Ising system of spins $S_i=\pm 1$ with Hamiltonian:
\be
\bt \hs = -\spv (J_{ij}S_iS_j-|J_{ij}|)
\lb{Ham}
\ee
where $\si$ is the spin configuration, the sum is over all interacting 
spin pairs and, as usually, $\bt=1/k_BT$. The constants in the Hamiltonian 
have been chosen for future convenience and for simplicity we can consider 
$J_{ij}=J>0$ $\forall i$,$j$, i.e. we take an Ising isotropic ferromagnet. 

The cluster formalism to describe ``droplets" of Ising spin
\cite{droplets} was originally developed by Kasteleyn and Fortuin 
\cite{KF}, and later on in a different approach by Coniglio and Klein 
\cite{CK}.
It is based on the mapping of the original model described by 
Hamiltonian (\ref{Ham}) into a new model in which spin couplings have 
infinite or zero strength. The mapping consists in stochastically changing 
the interactions $J_{ij}$ between spin pairs to new values $J'_{ij}$ 
(to define clusters we are interested in the limit 
$|J'_{ij}|\rightarrow\infty$ or $J'_{ij}\rightarrow 0$), in 
such a way that the two models, with new and old interactions, 
are statistically equivalent. 

To this aim, following KF, let's 
consider then each single couple of interacting spin $S_i$ and $S_j$, 
and suppose $J'_{ij}\in \{ 0, J', -J' \}$ with $J'>0$. 
If the change of $J_{ij}$ to a new interaction $J'_{ij}=J'$ 
occurs with a weight $p_0$, to a $J'_{ij}=-J'$ with $p_{\pi}$ and to 
$J'_{ij}=0$ with $q$, then the sum of statistical weights of a spin 
configuration $\{S_i,S_j\}$ in the new possible models is
\cite{KF,CK}:
\be
W(S_i,S_j)=q+p_0e^{J'(S_iS_j-1)}+p_{\pi}e^{-J'(S_iS_j+1)}
\lb{weight_S1S2_dilute}
\ee
It is to be noticed that we are slightly 
modifying KF-CK original approach in which $p_{\pi}$ was {\em a priori}
set to zero. 
To impose the statistical equivalence of the original and mapping models we 
must then require that a given spin configuration has the same weight:
\be
\exp(-\beta\hs)=\prod_{<i,j>} W(S_i,S_j)
\ee
So for each pair of interacting spins we must require:
\be
e^{J(S_iS_j-1)}=q+p_0e^{J'(S_iS_j-1)}+p_{\pi}e^{-J'(S_iS_j+1)}
\lb{KFeq}
\ee
where the $p_0$, $p_{\pi}$ and $q$ are unknown temperature functions. 

To introduce the definition of clusters of spins we must consider the 
limit $J'\rightarrow\infty$. 
Two spins $S_i$ and $S_j$ connected in the new model by an infinite 
interaction must have then a definite reciprocal direction 
(i.e. parallel if $J'_{ij}=+J'$ and antiparallel if $J'_{ij}=-J'$)
to have a non zero weight, 
otherwise, if disconnected, they are completely independent. Thus in the new 
model the {\em clusters} are naturally defined as the maximal sets of spins 
connected by $\infty$ interactions (called {\em bonds}).
The {\em deletion} ($J'_{ij}=0$) or the {\em freezing} 
($|J'_{ij}|=\infty$) of the original interactions leads to the contraction
of the spin lattice in independent fundamental units: the clusters.
In the limit $J'\rightarrow\infty$, eq.(\ref{KFeq}) becomes:
\be
e^{J(S_iS_j-1)}=q+p_0\delta_{S_i,S_j}+p_{\pi}\delta_{S_i,-S_j}
\lb{KFeqi}
\ee
This is a linear system of two equations with three unknowns 
$q$, $p_0$ and $p_{\pi}$, and so it is possible to 
introduce physical constraints 
to select some definite solution \cite{noistessi}. To this aim let's 
introduce the connectivity function $\gamma_{ij}$ , which is one if spin 
$S_i$ and $S_j$ belong to the same cluster and zero otherwise. 
It is possible to show that connectivity is always greater or equal to spin 
correlation \cite{noistessi,CLMP}:
\be
\lan S_iS_j \ran =\lan \gammaijpar\ran -\lan\gammaijnotpar\ran \leq 
\lan\gammaijpar\ran+\lan\gammaijnotpar\ran \equiv \lan \gamma_{ij} \ran
\lb{corconisi}
\ee
where $\gammaijpar$ ($\gammaijnotpar$) is one if $S_i$ and $S_j$ belong to 
the same cluster and are parallel (antiparallel).
A criterion which proved to be extremely important to select the definitions 
of interesting clusters 
(i.e. the relative value of $q$, $p_0$ and $p_{\pi}$) is to make 
connectivity as close as possible to correlation, i.e. to minimize 
connectivity as a function of $q$, $p_0$ and $p_{\pi}$
\cite{noistessi,artic1}:
\be
\lan \gamma_{ij} \ran \rightarrow minimum
\lb{conminisi}
\ee
This natural condition, which 
essentially corresponds to select clusters whose structure resembles the 
correlation patterns in the system, has given excellent results in 
frustrated and unfrustrated Ising spin systems \cite{noistessi,artic1}.
In the case of the simplest approximation in which we consider just a single 
couple of interacting spin $S_i$ and $S_j$, the mean connectivity is 
$\lan \gamma_{ij}\ran=(p_0+p_{\pi})/(1+e^{-2J})$, and to 
impose condition (\ref{conminisi}), with constraints 
$0\leq q,p_0,p_{\pi}$, naturally leads to $p_{\pi}=0$ (or 
analogously $p_0=0$ if $J<0$). In the present case, 
this results may be also simply obtained by directly imposing 
$\lan S_iS_j \ran = \lan \gamma_{ij} \ran$. 
The solution:
\be
\begin{array}{ccc}
p_{\pi}=0  & ; & p_0=1-q=1-e^{-2J}
\end{array}
\lb{solKF}
\ee
is the well known result by 
Kasteleyn-Fortuin \cite{KF} and by Coniglio-Klein \cite{CK}. 
Within this context it is possible to show that the partition function of a 
Q-Potts model \cite{Wu} may be written as 
($Q=2$ corresponds to Ising model) \cite{KF,CK}: 
\be
Z_Q(J)=\sum_C q^{|A|}p_0^{|C|}Q^{N(C)}
\lb{Z}
\ee
where $p_0=1-e^{-QJ}=1-q$, $N(C)$ is the number of clusters in the bonds 
configuration $C$ (i.e. the set of $\infty$ interactions), $|C|$ (resp.
$|A|$) is the total number of bonds in $C$ (resp. of absent bonds or 
zero interactions), and $\sum_C$ is the sum over all bonds configurations.
Eq.(\ref{Z}) gives the Ising partition function in terms of the partition 
function of a correlated-percolation model \cite{staufferbook}. 
Moreover, with KF solution, for an Ising ferromagnet, eq.(\ref{corconisi}) 
becomes:
\be
\lan S_iS_j \ran =\lan \gamma_{ij} \ran
\lb{corconisi1}
\ee

\section{Cluster formalism for $XY$ spin Hamiltonians}

Let's examine now the problem of cluster definitions in 
continuous spin systems. We consider an $XY$ model, but the same arguments may be 
extended to $O(n)$ models. Specifically, we consider a system of planar 
spins, $S_i$, with pair Hamiltonian:
\be
\bt H_{ij} = -(J_{ij}\cos(\theta_i-\theta_j)-|J_{ij}|)
\lb{Hamxy}
\ee
where $\theta_i$ is the phase of spin $S_i$, as above we suppose $\bt=1/k_BT$ 
and for clarity $J_{ij}=J>0$. The 
constant in the Hamiltonian has been chosen for convenience so that 
two ferromagnetically interacting spin have zero energy when 
they are parallel. 

Following the idea proposed by KF, we map the original model described by 
Hamiltonian (\ref{Hamxy}) into a new model in which the pair Hamiltonian 
between interacting spins is stochastically changed to new functional 
values $H'_{ij}$, in such a way that the two models are statistically 
equivalent. As above, to individuate clusters we are interested in 
the limit $H'_{ij}\rightarrow\infty$ or $H'_{ij}\rightarrow 0$. 
The main difference with the previous section will consists in the fact 
that many choices for $H'$ are necessary, but the arguments will be the 
same. Let's so define new pair Hamiltonians characterized by a new 
variable $\phi'_{ij}$:
\be
\bt H(S_i,S_j;\phi'_{ij})
\equiv  -(J'_{ij}\cos(\theta_i-\theta_j-\phi'_{ij})-|J'_{ij}|)
+C(J'_{ij})
\lb{Hamxy1}
\ee
where the parameter is $\phi'_{ij}\in[ 0,2\pi]$, $J'_{ij}=J'>0$ and $C(J')$ is 
an adjustable regularization function for the limit $J'\rightarrow\infty$.

To impose the statistical equivalence of the original and the mapping 
models, we then require that spin configurations have the same weight 
in both of them, so
if we define $p(\phi')$ as the statistical weight to map the pair Hamiltonian 
(\ref{Hamxy}) into $H(S_i,S_j;\phi')$, and $q$ the weight to map it into a 
zero energy interaction, the equation corresponding to the (\ref{KFeq}) 
becomes:
\be
e^{-\bt H_{ij}(S_i,S_j)}=q+
\int_{0}^{2\pi}p(\phi')e^{-\bt H(S_i,S_j;\phi')} d\phi'
\lb{KFeqxy}
\ee

As above, to define clusters we consider the limit $J'\rightarrow\infty$. 
Two spins $S_i$ and $S_j$ connected in the new model $H(S_i,S_j;\phi'_{ij})$ 
by an infinite interaction, must have a definite reciprocal direction 
(i.e. $\theta_i-\theta_j=\phi'_{ij}$) to have a non zero weight, 
otherwise, if not connected, they are completely independent. Thus in the new 
model the {\em clusters} are naturally defined as the maximal sets of spins 
connected by $\infty$ interactions ({\em bonds}). In contrast to the Ising 
case we now have much more than just two kind of bonds (between
parallel, $p_0$, 
or antiparallel spins, $p_\pi$). Following this method 
it is then possible to generalize the procedure of {\em deletion} 
($J'_{ij}=0$) and {\em freezing} ($J'_{ij}=\infty$) of the original 
interactions. Also in this case different clusters are independent (if $q$ 
is a function of $\theta_i-\theta_j$ or if the infinite limit is not 
systematically taken, then one has interacting clusters).
In the infinite limit $J'\rightarrow\infty$, eq.(\ref{KFeqxy}) becomes:
\be
e^{-\bt H_{ij}(S_i,S_j)}=q+
\int_{0}^{2\pi}p(\phi')\delta(\theta_i-\theta_j-\phi') d\phi'=
q+p(\theta_i-\theta_j)
\lb{KFeqxyi}
\ee
where the function $C(J')$ has been absorbed to regularize the definition 
of the $\delta$-function in the interval $[0,2\pi]$ with argument 
defined modulus $2\pi$. 
Eqs.(\ref{KFeqxyi}) is a linear functional equations in the unknown functions 
$p(\phi',J_{ij})$ and $q(J_{ij})$. 

A solution of eq.(\ref{KFeqxyi}) is suggested by the reasonable limit behavior 
$p(\phi')\rightarrow 0$ if $J\rightarrow 0$, or alternatively by the 
condition of local minimal connectivity (see below):
\be
\begin{array}{ccc}
q=e^{-2J} & ; & p(\phi')=e^{J(cos(\phi')-1)}-q
\end{array}
\lb{solxy}
\ee
This solution reproduces in the Ising case the results by Kasteleyn and 
Fortuin given in eq.(\ref{solKF}). 

The clusters are operatively individuated by the conditioned probabilities:
\be
p(\phi'|\theta_i-\theta_j)=p(\phi')\delta(\theta_i-\theta_j-\phi')
 e^{\bt H(S_i,S_j)} ~ ; ~~~~
q(\theta_i-\theta_j)=q \cdot e^{\bt H(S_i,S_j)} ~ .
\lb{proconxyp}
\ee
which may be respectively interpreted as the conditioned probability to 
substitute the original interaction in the mapping model with a bond of 
the kind $\phi'$ or with a zero interaction, given the spin configuration 
$\{S_i,S_j\}$ (note that this probabilities are completely independent 
on the choice of the constants for the energy of the ground state).
These conditioned probabilities may be used to implement MC cluster algorithms 
because they contain the necessary information to build clusters from 
spin configurations and it may be proved that algorithms based on 
these probabilities satisfy detailed balance principle 
\cite{BADK,noistessi}. They are the generalization to $XY$ of KF 
bond conditioned probabilities in Ising systems. 

Eq.(\ref{KFeqxyi}) may also be considered directly as the starting point to 
define clusters of bonds variables $\{\phi'_{ij}\}$, avoiding at all to 
introduce the procedures of Hamiltonian mapping and definitions (\ref{Hamxy1}). 
In this perspective eq.(\ref{KFeqxyi}) is just a way to introduce a 
statistical systems 
of variables of spin and bonds $(\{\theta_i\},\{\phi'_{ij}\})$ with 
the following peculiar properties \cite{EdwardsandSokal}: 
the marginal distribution of 
the $\{\theta_i\}$ is exactly equal to the Boltzman weight 
$e^{-\bt H(\{\theta_i\})}$; 
the conditional distribution of the $\{\phi'_{ij}\}$, given the 
$\{\theta_i\}$, is exactly expressed by eqs.(\ref{proconxyp}); 
the conditional distribution of the $\{\theta_i\}$, given the 
$\{\phi'_{ij}\}$, correspond to the above given definition of clusters, i.e.  
two interacting spin $S_i$ and $S_j$ belonging to the same cluster 
must have the definite reciprocal direction $\theta_i-\theta_j=\phi'_{ij}$, 
otherwise, if disconnected, they are completely independent. Note that 
only the sets $\{\phi'_{ij}\}$ such that given any two sites $h$ and $k$ 
the quantity $\Delta\theta=\sum_h^k \phi'_{ij}$ is independent of the 
``integration" path, are allowed.

\section{Relations between thermodynamics and percolation}

The previous section was devoted to introduce a simple generalization
of Kasteleyn-Fortuin and Coniglio-Klein clusters in $XY$
models. 
Now we face the problem to work out some main 
relations between percolative and thermodynamic 
quantities. 
Easy extensions may be given for general $O(n)$ models.

The partition function of the $XY$ model,
from eq.(\ref{KFeqxyi}), may be written as:
\be
Z\equiv \prod_i\int_0^{2\pi}d\theta_i e^{-\bt \sum_{<i,j>} H(S_i,S_j)}=
\sum_{\tilde C}^* q^{|A|}P(\tilde C)(2\pi)^{N(\tilde C)}
\lb{Zxy}
\ee                                               
where 
$|A|$ is the number of 
absent bonds on the lattice fixed the bonds configuration $\tilde C$, 
$N(\tilde C)$ is the total number of clusters in $\tilde C$, and 
by definition $P(\tilde C)=\prod_{<ij>\in \tilde C} 
p(\phi'_{ij})$. The sum, $\sum_{\tilde C}^*$, 
is intended over all possible bonds configurations, $\tilde C$, 
(note that two bonds configurations are distinguished by their geometry and by 
the kind of bonds $\{\phi'_{ij}\}$ they have), and specifically: 
\be
\sum_{\tilde C}^*\equiv 
\sum_{\tilde C}\int (\prod_{<ij>\in \tilde C} d\phi'_{ij}) ~ 
\delta(\tilde C,\{\phi'_{ij}\})
\ee
with $\prod_{<ij>\in \tilde C}$ the product over all bonds present in the 
configuration $\tilde C$, $\{\phi'_{ij}\}$ the set of indexes of such 
present bonds
and $\delta(\tilde C,\{\phi'_{ij}\})$ a function nonzero only if the 
configuration 
$\tilde C$ and the set $\{\phi'_{ij}\}$ are compatible 
(i.e. if the sum of $\phi'_{ij}$ between whatever 
fixed extrema $h$ and $k$ along a chain of present bonds, is independent of 
the path, i.e. clusters are well defined because such quantity is 
exactly the phase difference between $S_h$ and $S_k$). 

The percolative quantity to be compared to the 
thermodynamic two point correlation function is the pair connectivity 
$c(i,j)$, defined as:
\be
c(i,j)=\int_0^{2\pi} c(i,j,\phi) d\phi
\lb{connxy}
\ee
where $c(i,j,\phi)=\lan \gamma_{ij}(\phi) \ran$ and 
$\gamma_{ij}(\phi)=\gamma_{ij} \cdot \delta(\theta_i-\theta_j-\phi)$ which 
is zero if spin $S_i$ and $S_j$ do not belong to the same cluster 
or have a phase difference $\theta_i-\theta_j \neq \phi$. 
$c(i,j,\phi)$ is so the probability of spin $i$ and $j$ to 
belong to the same cluster with a phase difference $\phi$. 

It is possible to show that the pair correlation function 
$g(i,j)= \lan S_i\cdot S_j \ran $ is given by:
\be
g(i,j)=\int_0^{2\pi} \cos(\phi) c(i,j,\phi) d\phi 
\lb{corrxy}
\ee
Eqs.(\ref{connxy}) and (\ref{corrxy}) imply $g(i,j) \leq c(i,j)$, analogously 
to Ising systems where eq.~(\ref{corconisi}) holds. 
A consequence of this proven inequality is that 
$T_c\leq T_p$, where $T_c$ and $T_p$ are defined as the temperatures where 
respectively the magnetic susceptivity $\chi$ and the mean cluster size 
$S=\sum_{s}^{'} n_s s^2$ \cite{staufferbook}
($n_s$ is the number of clusters of size $s$ and 
$\sum_{s}^{'}$ is the sum over just finite clusters) become 
singular. These relations naturally suggest, in 
analogy with the Ising spin case, the criterion of minimal connectivity to 
select clusters definitions, i.e. to impose the condition:
\be
c(i,j)\rightarrow minimum
\lb{connminxy}
\ee
where $c(i,j)$ has to be minimized respect to $p(\phi)$.
As anticipated above, imposing eq.(\ref{connminxy}) just 
for each single couple of interacting spin $S_i$ and $S_j$, 
directly leads to select the solution given in eq.(\ref{solxy}) for 
eq.(\ref{KFeqxyi}). It is to be noticed that this solution is however
just the simplest extensions of KF result. In facts more general solutions 
must be found as showed below, but the general tools introduced
to link percolative and spin properties, allows to exploit, for $XY$
models, the many 
techniques to individuate and manipulate clusters known in the literature 
for Ising systems (see \cite{KF,CK,noistessi,artic1}). In what follows 
we will restrict however to consider the simple solution given in 
eq.(\ref{solxy}).

The relations above reported indicate that thermodynamic spin quantities
may be generally expressed in terms of cluster properties. 
For example it is 
possible to link the mean energy $E$ with geometrical quantities. 
In the case of an isotropic ferromagnetic $XY$ model 
it results: 
\be
E/N=-\lan\gamma_{0 1}\partial_{\bt}\ln(p(\phi'_{0 1})/q)\ran
+\partial_{\bt}\ln(q)
\lb{Exy}
\ee
where $0$ and $1$ are two of the $N$ interacting pairs of spin in the 
system. As eq.(\ref{Zxy}) is the natural generalization of eq.(\ref{Z}), 
so eq.(\ref{Exy}) is the extension of the corresponding energy-bond relation 
in Ising systems \cite{noistessi,Hu}. 

\section{Clusters in one dimensional $XY$ model}

To understand the properties of the above defined clusters it may be 
interesting to analyze the question in some details.
In a one dimensional $XY$ model of nearest neighbor interacting spin
\cite{Stanleybook}, the 
geometry of the above defined clusters corresponds to 
chains of bonds, and the problem is extremely simplified. So it is 
possible to prove that, adopting solution (\ref{solxy}), the partition 
function, eq.(\ref{Zxy}), in the 
case of an isotropic $XY$ ferromagnetic chain, is:
\be
Z_{1D}(J)=\sum_C q^{|A|}P^{|C|}(2\pi)^{N(C)}
\lb{Zxy1D}
\ee
where $\sum_C$ is just the sum over all graphs of bonds on the chain, 
$P=2\pi e^{-J} (I_0(J)-e^{-J})$ ($I_n(x)$ is the imaginary argument 
Bessel function of order $n$).
From eq.(\ref{Zxy1D}) it is possible to see that the 
partition function of an $XY$ chain may be written as that of a Q-Potts 
linear model, $Z_Q$ (see eq.(\ref{Z})), times a simple factor. Specifically:
\be
Z_{1D}(J)=K \cdot Z_Q(J_Q) 
\lb{Z1Dezqp}
\ee
where $N$ is the total number of interactions, $Q=2\pi$, 
$Q \cdot J_Q(J)=\ln(1+P/q)$ ($J_Q\sim J$ if $J\rightarrow \infty$ or 
$J\rightarrow 0$), and $K(N,J)=e^{-N(2J-Q J_Q)}$. 

We are concerned with clusters 
and spin properties, and in the context of the linear model it is possible to 
prove a definite 
relation between correlation $g$ and connectivity $c$:
\be
g(i,j)=e^{-r/\xi} c(i,j)
\lb{conncorr1D}
\ee
where $r=|i-j|$ is the number of spin between $S_i$ and $S_j$ plus one, and 
\be
\xi^{-1}(J)=\ln[(I_0(J)-e^{-J})/I_1(J)] 
\lb{chsixy1D}
\ee
Moreover the connectivity $c(i,j)$ at temperature $T=1/J$ for the $XY$ model, 
is equal to the connectivity $c_Q(i,j)$ of the KF (or CK) clusters introduced 
above in the Q-Potts model with $T=1/J_Q(J)$ and $Q=2\pi$:
\be
c(i,j)|_J=c_Q(i,j)|_{J_Q(J)} 
\lb{conn1Dq}
\ee
These results should shed some light on 
the connections between cluster connectivity and spin 
correlation in one dimensional $XY$ models. A trivial consequence of all 
these relations in 1D is that $T_p=T_c=0$ (it would be hard to find 
clusters with a bit of randomness with $T_p>0$ in one dimension), but this 
coincidence does not hold for the critical behavior.
Defining $\xi_{XY}$ as the correlation length in the $XY$ model and 
$\xi_{Q-Potts}$ as the mean cluster radius in its Q-Potts equivalent, 
eq.(\ref{conncorr1D}) imposes $\xi_{XY}^{-1}=\xi_{Q-Potts}^{-1}+\xi^{-1}$, 
but, in 1D, at low temperature $\xi_{XY}\sim T^{-1}$ while 
$\xi_{Q-Potts}\sim e^{A(Q)/T}$ \cite{staufferbook,Stanleybook}, and so 
clusters quite loosely express spin-spin correlations.

\section{MC results for $XY$ model in higher dimensions}

The analytical problem concerning the structure of clusters in $XY$ models in 
higher dimensions, is, worthless to say, much more difficult. 
We present then some Monte Carlo results about clusters properties 
(defined from solution (\ref{solxy})), in two and three dimensions. 

MC simulations were done using a standard Metropolis 
spin flip algorithm \cite{Binder}
on $n$-clock models on a square or cubic 
lattice described by 
Hamiltonian (\ref{Hamxy}) (in the following if not specified we will 
consider the isotropic case $J_{ij}=J\geq 0$), whose spin 
$S_i$ have a phase $\theta_i=2\pi m/n$ with 
$m\in \{0,...,n-1 \}$.

Let's briefly examine our MC results in two dimension.
In the case $n=2$ we exactly recover the 2D ferromagnetic Ising model. Our 
MC simulation indicate the well known result of equal critical temperatures 
$T_c=T_p=2.269$  (all temperature are measured in unit of $J$) 
and a percolative critical phenomena characterized by Ising 
exponents. Moreover, the MC dynamic based on the above defined 
clusters is just the Swendsen-Wang dynamic and the phenomena of critical 
slowing down drastically reduced \cite{SW,Wolff}. 

For $n > 4$ such correspondence is no longer verified. Such a result may be 
expected because in these cases connectivity and correlation are 
not coinciding as shown by the simple example of just 
two interacting spin with $n=5$ (the cases $n=3,4$ may be successfully faced 
with some tricks, in resemblance of the possibility to map 
$n=3,4$ clock-models in an equivalent Q-Potts \cite{Wu}).

The percolation critical temperature decreases for increasing $n$, and  
approaches a plateau in the large $n$ limit. 
In actual facts, via MC simulations, for $n=36$, we find that the 
percolation point is at 
\begin{center}
$T_p=1.69\pm 0.03$
\end{center}
to be compared with the 2D $XY$ critical temperature at $T_c\sim 0.89$ 
\cite{Wolff,Henley}. 
We find that the percolation critical exponents are in the 
universality class of random percolation as expected 
because there is no thermodynamic transition underlying the percolative 
one: the critical exponents, measured via a finite size scaling analysis 
\cite{Binder} (reported in Fig. 1), are $\nu=1.33\pm0.05$ and 
$\gamma/\nu=1.79\pm0.05$ in perfect agreement with 2D random percolation 
exact values $\nu=4/3$ and $\gamma/\nu=43/24$. 
These exponents, in percolation, characterize respectively 
the divergence of mean cluster radius $\xi$ and mean cluster size $S$: 
$\xi\sim |T-T_p|^{-\nu}$ and $S\sim |T-T_p|^{-\gamma}$ 
\cite{staufferbook}. 

This behavior is observed, as may be easily suspected, in 
frustrated or disordered systems too. 
We tested, via MC, the Fully Frustrated $XY$ model (FF) and the 
$\pm J$ $XY$ Spin Glass (SG) (see references in \cite{Jose-Ramirez}), 
where we found the same 
percolative critical exponents and (see the scaling analysis in Fig. 2 and 3):
\begin{center}
$T^{FF}_p=1.61\pm 0.03$ and $T^{SG}_p=1.64\pm 0.03$
\end{center}
This value for the percolation transition in the FF model is above the 
critical region located around $T\sim 0.4 \div 0.5$ (see 
\cite{Jose-Ramirez}). Also Wolff's clusters show a percolation point well 
above the critical region \cite{Henley}, but it is possible to 
introduce their direct generalizations whose $T_p$ may be pushed closer 
and closer to it \cite{Cataudellaxy}.

The same kind of results are found in three dimension. 
For $n=2$ we recover the well known properties of KF or SW clusters in the 
3D Ising model 
$T_p=T_c\sim 4.5$ and $\nu\sim 0.62$, $\gamma/\nu\sim 1.97$ (see 
\cite{FerrenbergLandau}).
Our MC runs show, for $n=36$, 
$T_p=3.75\pm 0.05$ 
and 3D random percolation critical behavior with $\nu=0.87\pm .05$ and 
$\gamma/\nu=2.00\pm .05$ (see Fig.4). 
These values are to be compared to the results of 3D $XY$ $T_c\sim 2.2$ 
and $\nu\sim 0.66$, $\gamma/\nu\sim 1.98$ (see \cite{Gottlob}). 
Essentially the same values are found for the 3D 
$\pm J$ $XY$ Spin Glass. 

It is interesting to note that, as expected and discussed above
(see also \cite{noistessi,artic1}), whenever the gap between $T_p$ and $T_c$ 
becomes finite, SW like cluster algorithms for MC 
simulations become unable to reduce critical slowing down.

After these MC results, 
the panorama we get illustrates that the straight generalization to 
$XY$ models of KF clusters, given in eq.(\ref{solxy}), has not the peculiar 
properties of KF clusters in Ising like systems: the thermodynamic and 
percolative transition are no longer coincident. 
In Ising models more complex procedures have been introduced to
individuate physical clusters, as those proposed in 
\cite{BADK,noistessi,artic1}.
It would be interesting to verify if the 
extensions to $XY$ of such procedures according the lines 
proposed above, have the same percolation properties here found or 
new interesting results can be obtained. 

\section{Summary and conclusions}

In analogy to Kasteleyn and 
Fortuin and Coniglio and Klein works, 
cluster of nearest neighbor spin in $XY$ models may be defined 
as the sets of spin connected by bonds according definite rules.
The clusters divide the original lattice into independent regions of 
statistically coherent spins.
Kasteleyn and Fortuin percolative concepts and tools to link clusters 
and spin properties, which proved to be so useful in Ising systems, 
can be so extended to $XY$ models. 
In these models, at a first simple level, KF clusters may be defined
by separately looking at just each single couple of interacting spins. 
Consequently, bonds are introduced between them according 
the definite probability distribution given in equation (\ref{solxy}) 
(as a matter of fact, this is for $XY$ models, but the analog for 
general $O(n)$ is absolutely similar). 

Nevertheless, at the simple level here explored, 
many differences appear with Ising systems. 
It is known that in the Ising model these clusters have a percolation 
point which, imposing condition (\ref{connminxy}), may be pushed to 
coincide with the critical one $T_c$, and has percolative exponents 
in the Ising universality class \cite{KF,CK}.
The properties of the new clusters may be studied analytically 
for continuous spin systems in one dimension. Here  
eq.(\ref{conncorr1D}) implies that clusters mean square 
radius, differently from Ising KF case, is no longer coincident with 
the correlation length in the system. 
Numerical results show the same behavior in higher dimensionality, 
where moreover cluster percolation point $T_p$ is different from $T_c$. 
Phenomenologically, the temperature 
$T_p$ is the point where regions of almost parallel 
spin (in ferromagnetic models), i.e. regions of coherent spins,  
percolate in the system. 
The transition corresponding to this point, for clusters defined by 
eq.(\ref{solxy}), is in the random percolation universality class.

The occurrence of a finite gap between $T_p$ and $T_c$ is found in 
Ising spin systems when frustration is present. In this cases 
a general criterion to close such a gap has been proposed 
\cite{BADK,noistessi,artic1}. Exploiting the results here presented, it 
is possible to apply such a criterion to frustrated and unfrustrated 
$O(n)$ models too, and in perspective give a percolative description 
of their critical behaviors in analogy to the known results for 
Ising like models. This approach would lead to a change of 
the bond probability distribution given in eq.(\ref{solxy}). 
In unfrustrated and frustrated Ising models a definite physical 
origin has been associated to the percolation point $T_p$ \cite{conglasses}. 
It is then natural to speculate on it in continuous spin models too.

The criterion introduced in \cite{BADK,noistessi,artic1} is actually suited 
to develop efficient MC cluster algorithms in Ising systems. 
The perspective to 
go further in such a direction also for $O(n)$ models, is very appealing.

\bigskip

The author is grateful to Prof. Antonio Coniglio for stimulating 
discussions and suggestions.

\section*{Figure Captions.}

\begin{description}

\item{Fig. 1.} Finite size scaling of the mean cluster size $S$ 
of clusters defined by bond probabilities given in eq.(\ref{proconxyp}), 
in the 2D $XY$ ferromagnet. The scaling parameters are 
$T_p=1.69\pm .03$ and $\nu=1.33\pm .05$ $\gamma/\nu=1.79 \pm .05$

\item{Fig. 2.} Finite size scaling of the mean cluster size $S$ 
of clusters defined by bond probabilities given in eq.(\ref{proconxyp}),
in the 2D $XY$ Fully Frustrated. The scaling parameters are
$T_p=1.61\pm .03$ and $\nu=1.33\pm .05$ $\gamma/\nu=1.79 \pm .05$

\item{Fig. 3.} Finite size scaling of the mean cluster size $S$ 
of clusters defined by bond probabilities given in eq.(\ref{proconxyp}),
in the 2D $XY$ $\pm J$ Spin Glass. The scaling parameters are
$T_p=1.64\pm .03$ and $\nu=1.33\pm .05$ $\gamma/\nu=1.79 \pm .05$

\item{Fig. 4.} Finite size scaling of the mean cluster size $S$ 
of clusters defined by bond probabilities given in eq.(\ref{proconxyp}),
in the 3D $XY$ ferromagnet. The scaling parameters are
$T_p=3.75\pm .05$ and $\nu=0.87\pm .05$ $\gamma/\nu=2.00 \pm .05$

\end{description}


\begin{thebibliography}{9999}

\bibitem{KF} 
 P. W. Kasteleyn and C. M. Fortuin, J. Phys. Soc. Japan Suppl. {\bf 26}, 11
 (1969);
 C. M. Fortuin and P. W. Kasteleyn, Physica (Utrecht) {\bf 57}, 536 (1972).

\bibitem{CK} A. Coniglio and W. Klein, J. Phys. A {\bf 13}, 2775 (1980).

\bibitem{droplets} M. E. Fisher, Physics (N.Y.) {\bf 3}, 225 (1967). 

\bibitem{Wu} F. Wu, Rev. Mod. Phys., {\bf 54}, 235 (1982).

\bibitem{89Co} A. Coniglio, Phys. Rev. Lett., {\bf62 } 3054 (1989); 
A. Coniglio in {\em Correlation and connettivity-Geometric 
 aspects of Physics,Chemistry and Biology} NATO ASI series vol. 188 (1990), 
 ed. E. Stanley, W. Ostrowsky.

\bibitem{SW}  
 R. H. Swendsen and J. S. Wang, Phys. Rev. Lett., {\bf 58}, 86 (1987);
 J. S. Wang and R. Swendsen, Physica A, {\bf 167}, 565 (1990).
 
\bibitem{Wolff} U. Wolff, Phys. Rev. Lett., {\bf 60}, 1461 (1988).
 U. Wolff, Phys. Rev. Lett., {\bf 62}, 361 (1989).

\bibitem{BADK} 
 D. Kandel, R. Ben-Av and E. Domany, Phys. Rev. Lett., {\bf 65}, 941 (1990);
 D. Kandel, R. Ben-Av and E. Domany, Phys. Rev. B, {\bf 45}, 4700 (1992). 
 D. Kandel and E. Domany, Phys. Rev. B, {\bf 43}, 8539 (1991).

\bibitem{noistessi}
 V. Cataudella, G. Franzese, M. Nicodemi, A. Scala and A. Coniglio
 Phys. Rev. Lett., {\bf 72} 1541 (1994); Il Nuovo Cimento D, {\bf 16} 1259 
(1995); Phys. Rev. E, {\bf 54}, 175 (1996).

\bibitem{Gubernatis} N. Kawashima and J.E. Gubernatis, Phys. Rev. E, {\bf 51},
  1547 (1995).

\bibitem{Henley} P.W. Leung and C.L. Henley, Phys. Rev. B, {\bf 43} 752 (1991).

\bibitem{CLMP} A. Coniglio, F. di Liberto, G. Monroy, F. Peruggi 
 Phys. Rev. B, {\bf 44} 12605 (1991).

\bibitem{staufferbook} D. Stauffer and A. Aharony, {\em Introduction to 
 percolation theory}, second edition (London: Taylor and Francis) 1995.

\bibitem{artic1} M. Nicodemi, J. Phys. A, {\bf 29}, {\em 1961 (1996)}.

\bibitem{EdwardsandSokal} R.G. Edwards and A.D. Sokal, 
 Phys. Rev. D, {\bf 38} 2009 (1988).

\bibitem{Hu} C.S. Hu and S.S. Hsiao, Phisica A, {\bf 184}, 192 (1992).

\bibitem{Stanleybook} 
 H.E. Stanley, {\em Phase Transitions and Critical Phenomena } 
(Clarendon Press, 1971). 

\bibitem{Binder} 
 K. Binder, D. W. Hermann, {\em Monte Carlo Simulation in
 Statistical Physics}, (Springer-Verlag, Berlin, 1988). 

\bibitem{Jose-Ramirez} J. Villain, J. Phys. C {\bf 10}, 
1771 and 4793 (1977). J. Lee, J.M. Kosterlitz and E. Granato, Phys. Rev. B 
{\bf 43}, 11531 (1992). G. Ramirez-Santiago and J.V. Jos\'e , 
Phys. Rev. E {\bf 49}, 9567 (1994). 
P. Olson, Phys. Rev. Lett. {\bf 75}, 2758 (1995).

\bibitem{Cataudellaxy} V. Cataudella and M. Nicodemi, to be published in 
Physica A.

\bibitem{FerrenbergLandau} A.M. Ferrenberg and D.P. Landau, Phys. Rev. B 
{\bf 44}, 5081 (1991).

\bibitem{Gottlob} A.P. Gottlob and M. Hasenbusch, Physica A {\bf 201}, 593 
(1993).

\bibitem{conglasses} see references in A. Coniglio, {\em Il Nuovo Cimento} 
{\bf 16D}, N.8 1027 (1994); and the proceedings of the Euroconference 
``Non equilibrium Phenomena in supercooled fluids, glasses and amorphous 
materials", Pisa (1995).

\end{thebibliography}
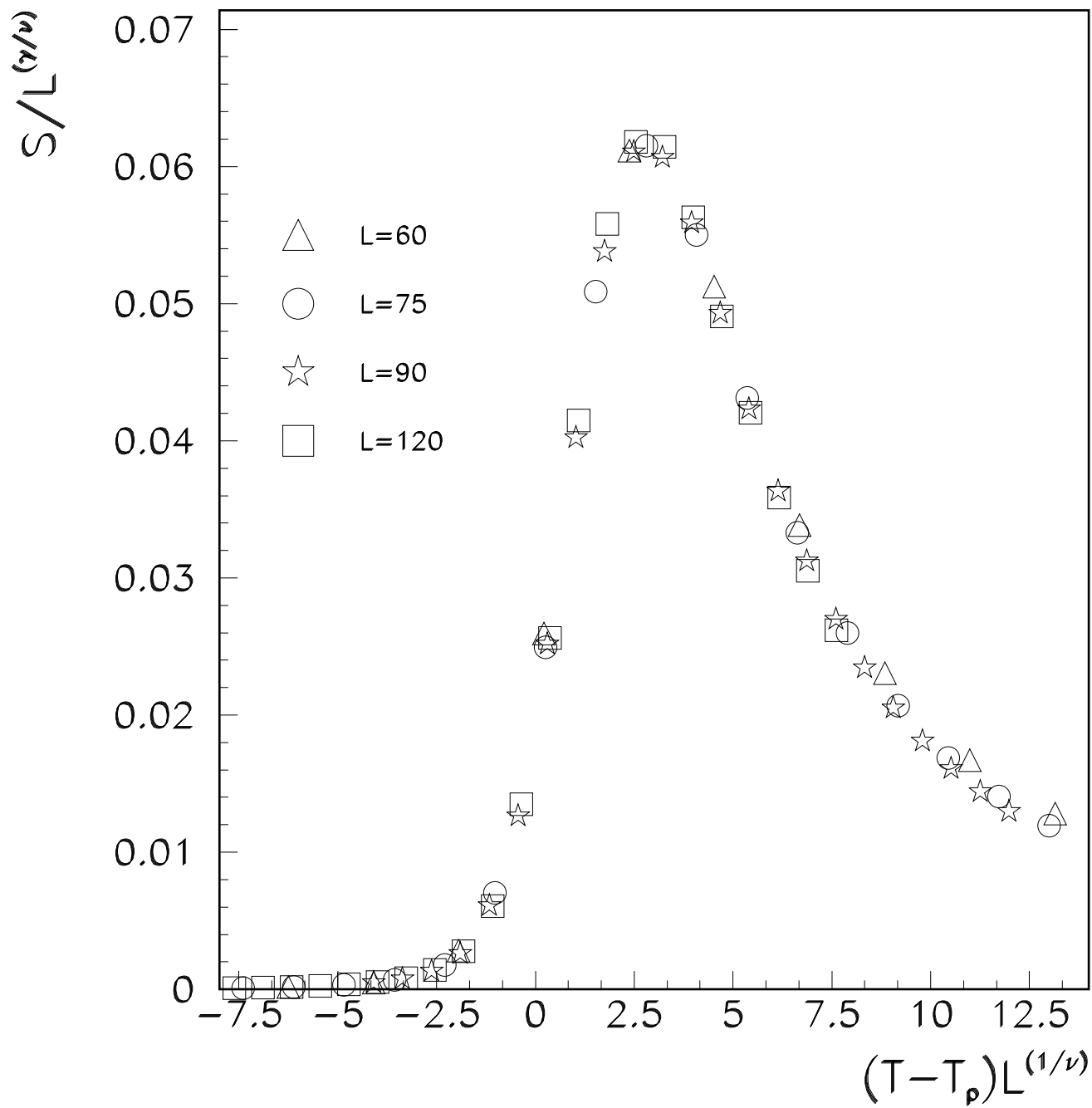
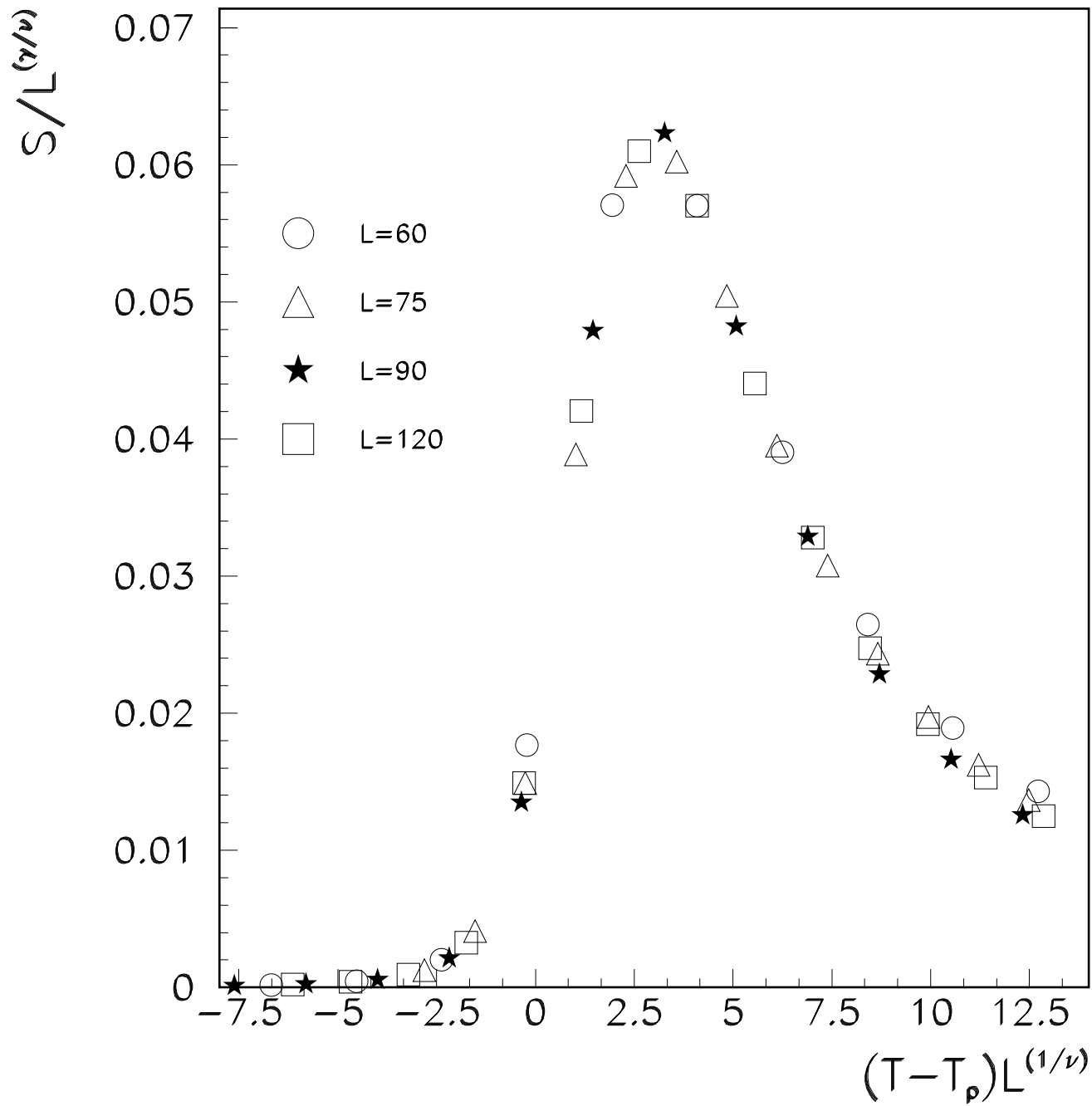
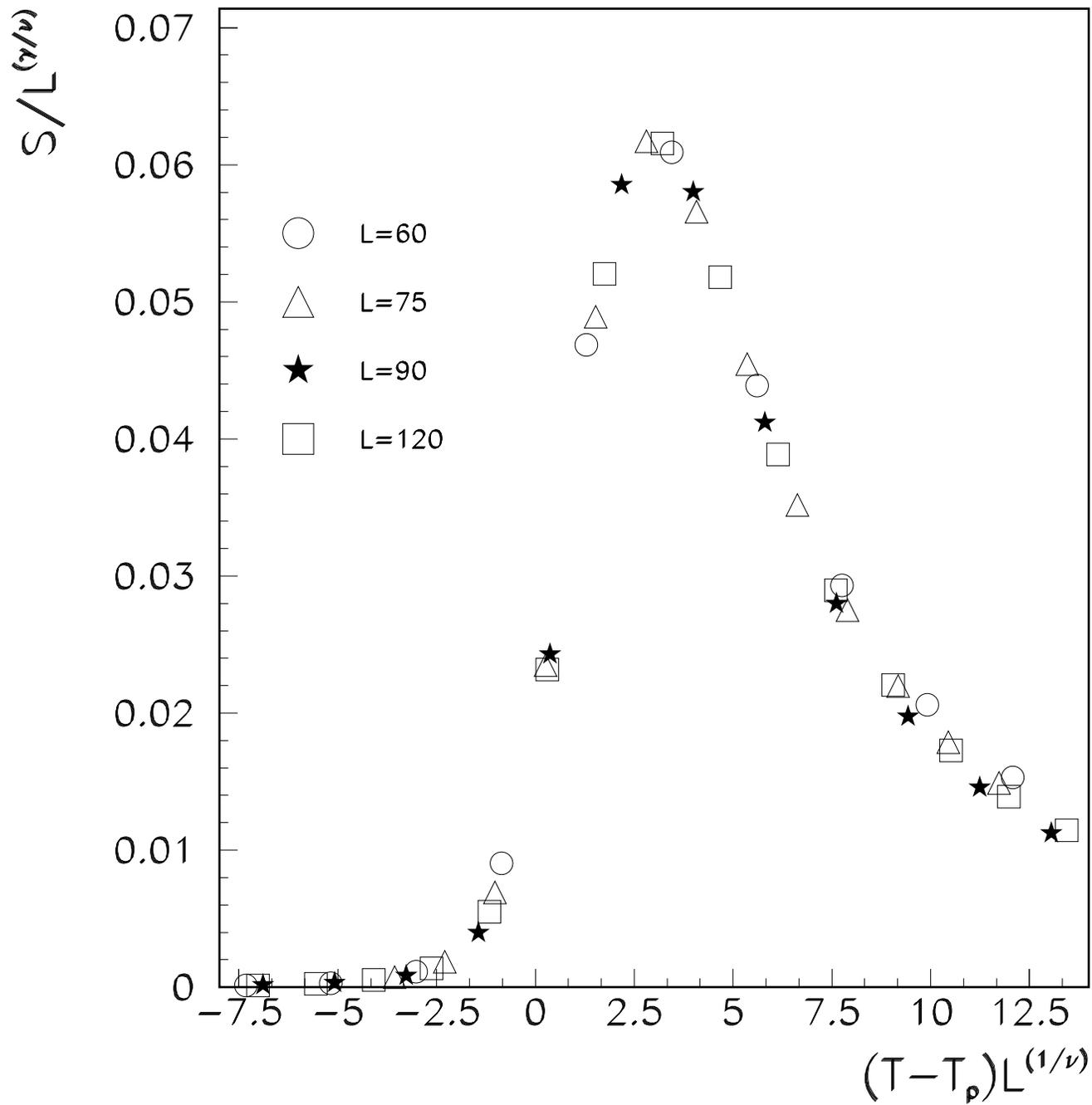
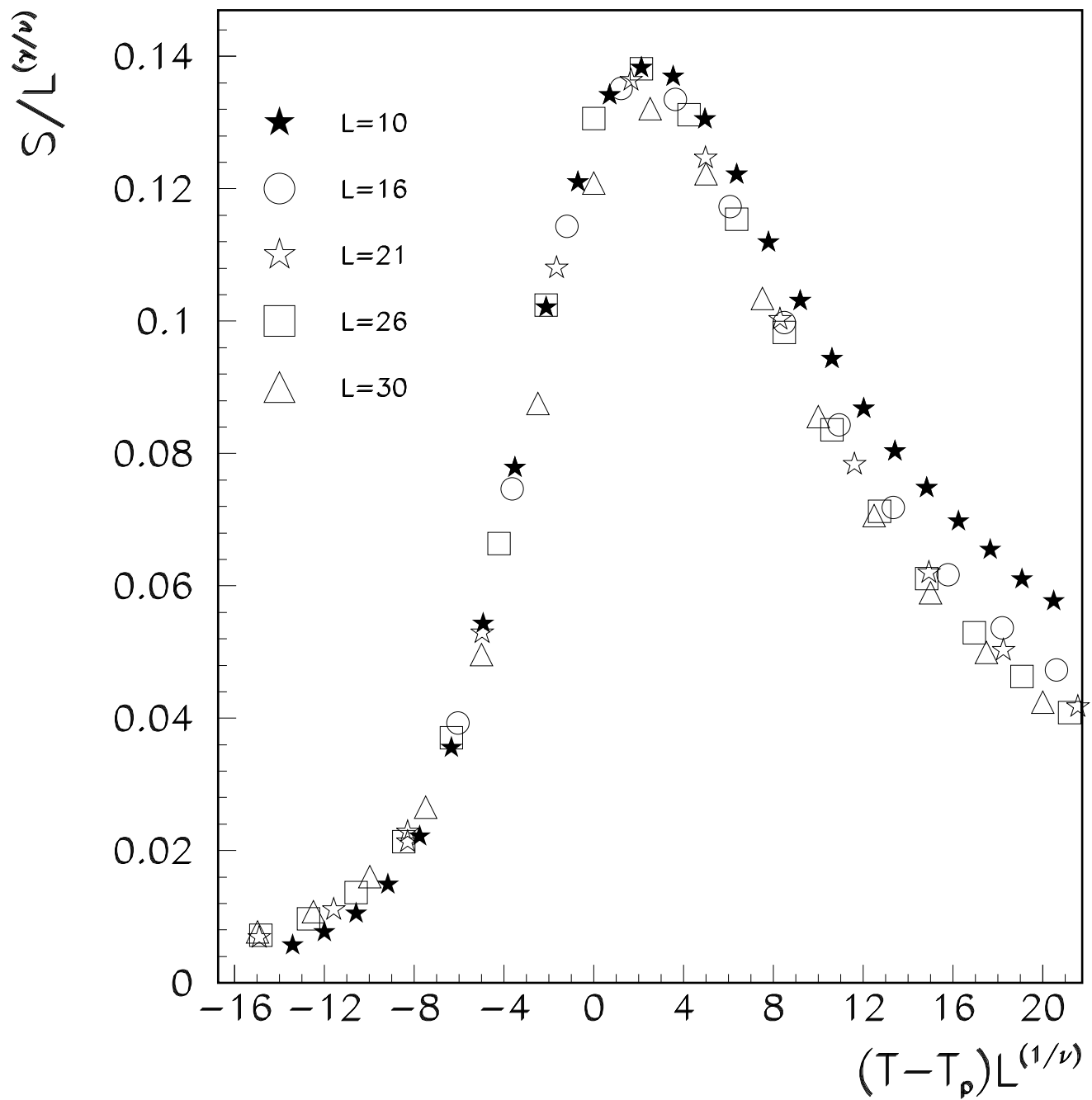
\end{document}